\documentclass[%
 aip,
 amsmath,amssymb,
 reprint,%
]{revtex4-1}

\usepackage[T1]{fontenc}
\usepackage[latin9]{inputenc}
\usepackage{bm}
\usepackage{bm}
\usepackage{amsmath}
\usepackage{amssymb}
\usepackage{graphicx}
\usepackage[normalem]{ulem}

\usepackage{color}

\begin{document}

\setlength{\arraycolsep}{2pt}
\title[Plasmon Bragg gratings]{Propagation of surface plasmons on plasmonic Bragg gratings }
%Scattering of a surface plasmon from a square-wave grating]{Scattering of a surface plasmon from a square-wave grating: An analytical approach}

\author{A. J. Chaves } 

\address{ Department of Physics, Instituto Tecnol\'ogico de Aeron\'autica,
DCTA, 12228-900 S\~ao Jos\'e dos Campos, Brazil}

\address{ Department and Centre of Physics, and QuantaLab, University of Minho,
Campus of Gualtar, 4710-057, Braga, Portugal} 

\author{N. M. R. Peres }

\address{ Department and Centre of Physics, and QuantaLab, University of Minho,
Campus of Gualtar, 4710-057, Braga, Portugal} 

\address{ International Iberian
Nanotechnology Laboratory (INL), Av. Mestre José Veiga, 4715-330 Braga,
Portugal}
 \email{peres@fisica.uminho.pt}
%\ead

\begin{abstract}
We use coupled-mode theory to describe the scattering of a surface-plasmon
polariton (SPP) from a square wave grating (Bragg grating) of finite
extension written on the surface of either a metal-dielectric interface
or a dielectric-dielectric interface covered with a patterned graphene sheet. We find analytical solutions for the
reflectance and transmittance of SPP's when only two modes (forward- and back-scattered) are considered. 
We show that in both cases the reflectance spectrum presents stop-bands where the
SPP is completely back-scattered, if the grating is not too shallow. In addition, the reflectance
coefficient shows Fabry-Pérot oscillations when the frequency of the
SPP is out of the stop-band region. For a single dielectric well, we show that
there are frequencies of transmission equal to 1. 
We also provide simple analytical expression for the different quantities in the electrostatic limit. 
\end{abstract}
\maketitle

\section{Introduction}

The usage of plasmonic technology depends on the possibility of controlling the propagation of surface-plasmon polaritons.
Bragg grattings are a relatively simple way to control the propagation of light in both optical fibers \cite{hill1997} and  metal-dielectric plasmonic interfaces\cite{han2007}. When the
conditions for destructive interference are fulfilled, the grating acts as a perfect mirror. In traditional Bragg gratings, the stop-band frequency can be engineered through the grating geometric parameters and a judiciously choice of 
dielectrics. For plasmonic graphene Bragg grattings, the stop-band depends on the carrier density (or the equivalently the Fermi energy of graphene) \cite{tao2014graphene}, thus
giving a new tool for  in-situ control of the stop-band through a gate potential. We  show in  this paper that coupled-mode theory can be used to obtain analytic expressions
describing the propagation of graphene plasmons as they propagate along the Bragg grating.

In a recent paper \cite{Feng} coupled-mode theory \cite{Huang-94} was used for describing
a set of experimental results showing unidirectional reflectionless
in parity-time metamaterial at optical frequencies (see also Ref. \cite{Ramezani-2011}). The metamaterial
was itself made of a periodic arrangement of metallic nanostructures in an optical fiber.
Coupled mode theory was quite popular in the seventies and the eighties of the last century and  to our best knowledge it was
first discussed by Yariv in the context of coupled waveguides \cite{Yariv-73}. However,
with the advent of powerful numerical methods its use has declined.
The paper of Feng \emph{et al}. \cite{Feng} gives a nice example
where coupled-mode theory allows an analytical analysis of a scattering
experiment with the obvious insight that an analytical solution provides
over a fully numerical one.

 In the context of guided wave optics, Taylor
and Yariv provided \cite{Taylor-74} a detailed analysis of co- and
contra-directional coupling, which corresponds to forward- and back-scattering
of a single propagating mode induced by a periodic perturbation.
Such perturbation can be introduced as a change of the dielectric function
along the propagation direction or  a Bragg grating
imposed on the surface of the waveguide. The theory of electromagnetic
propagation in periodic stratified media was first discussed by in
great detail by Yeh, Yariv, and Hong \cite{Yeh}. In the context of the theory of lasers, a comparison between the transfer matrix method and coupled-mode theory was given by Makino\cite{Makino-94}. Recently, coupled-mode theory was used for studying the scattering of  electromagnetic modes in a waveguide with corrugated boundaries \cite{Otto-2015}.

In this work, we discuss the application of coupled-mode theory to
the back-scattering of a surface-plasmon polariton from a one-dimensional
Bragg grating imposed on the surface of a metal-dielectric interface.
As a second example, we consider a graphene sheet covering a finite
Bragg grating and the back-scattering of a graphene
surface-plasmon polariton is discussed. To our best knowledge, coupled-mode
theory has not been applied so far to discuss the scattering of SPP's.
However, a recent work has used this approach to discuss the excitation of SPP in gratings by far-field coupling \cite{Lou-2016}. The coupling of a Gaussian laser beam to a SPP in a metallic film using coupled-mode theory was discussed by Ruan {\it et al.} \cite{Ruan-2014}. In the context of graphene physics, coupled-mode theory was recently used to discuss the excitation of localized plasmons  of a graphene-based cavity with a Silver waveguide \cite{Chen-2018}.

Also, problems in the context of nonlinear optics can be treated using coupled-mode theory \cite{Hong-2018}. Integrating numerically the coupled-mode equations, Petracek and Kuzmiak have described Anderson localization of channel SPP's in a disordered square-wave grating \cite{Petracek-2018}.
In a different context, Graczyk and Krawczyk have studied \cite{Graczyk-2017} the propagation of 
magnetoelastic waves using the methods described in this article.
The nonlinear interaction, promoted by a nonlinear second-order susceptibility tensor, between SPP's was considered first by 
Santamato and Maddalena
\cite{Santamato-1982}. Interesting enough, coupled-mode theory was adapted for describing coupling of Bose-Einstein condensates \cite{Kivshar-2000}.
The same type of approach has been used for describing
the field enhancement near plasmonic nanostructure under the effect of an external field \cite{Sun-2011}.
The extension of the theory to chiral waveguides was achieved by Pelet and Engheta \cite{Pelet-1990}, and constitutes a nice application of Lorentz's reciprocity theorem. % (see Appendix \ref{app:lorentz})
 
 Coupled-mode equations are able to give  both numeric and analytical results,that is,  they can either be numerically integrated, thus giving exact results, or  they can be solved in an approximate manner, thus giving approximate results. Both approaches have their own advantages. 
In this paper,
our analytical results are approximated in the sense that only two modes, forward-
and back-scattered modes, of the same frequency, are considered. 
This is a good approximation, since in the wave-guide only two SPP modes
exist (forward and backward propagating SPP modes). However
we do
neglect the possible emission of radiation when the SPP impinges on
the grating. Indeed, in the context of scattering of graphene's SPP's by abrupt
interfaces, it has been shown that the coupling of the SPP to the
radiation modes is weak \cite{Bludov,Chaves}  and, therefore, the same is expected here. In Ref. [\onlinecite{Chaves}], we 
have found that for graphene plasmons the losses due to radiative emission is proportional to the contrast of the refractive index $\epsilon_1-\epsilon_2$ (or the conductivities  $\sigma_1-\sigma_2 $), but even when $\sigma_2>2\sigma_1$, the losses due to the radiative emission were less
than 2\%. In these systems, the main mechanism for losses is the  intrinsic damping in the material.
We note, however, that
the formalism is general and there is no impediment to the inclusion
of both radiative and evanescent modes in it. The price to pay may
be the lack of an analytical solution.

The paper is organized as follows: in Sec. \ref{sec:simple} we simplify the coupled-mode equations, expressing them in terms of forward and back-scattered amplitudes.
In Sec. \ref{sec:exact_sol} we particularize coupled-mode equations to the case where only two degenerate modes in frequency are coupled and find a general solution using transfer-matrix method for a square-wave Bragg grating.
In Sec. \ref{sec:SPP} (Sec. \ref{sec:GSPP})  the scattering of metallic SPP's  (graphene SPP's) from a Bragg grating is discussed. We conclude the paper with a brief discussion in Sec. \ref{sec:conc}.

%Below we will consider $\epsilon_{3}=\epsilon_{1}=1$.  \label{fig:Unit-cell-of} }

\begin{figure}
\centering
\includegraphics[width=8cm]{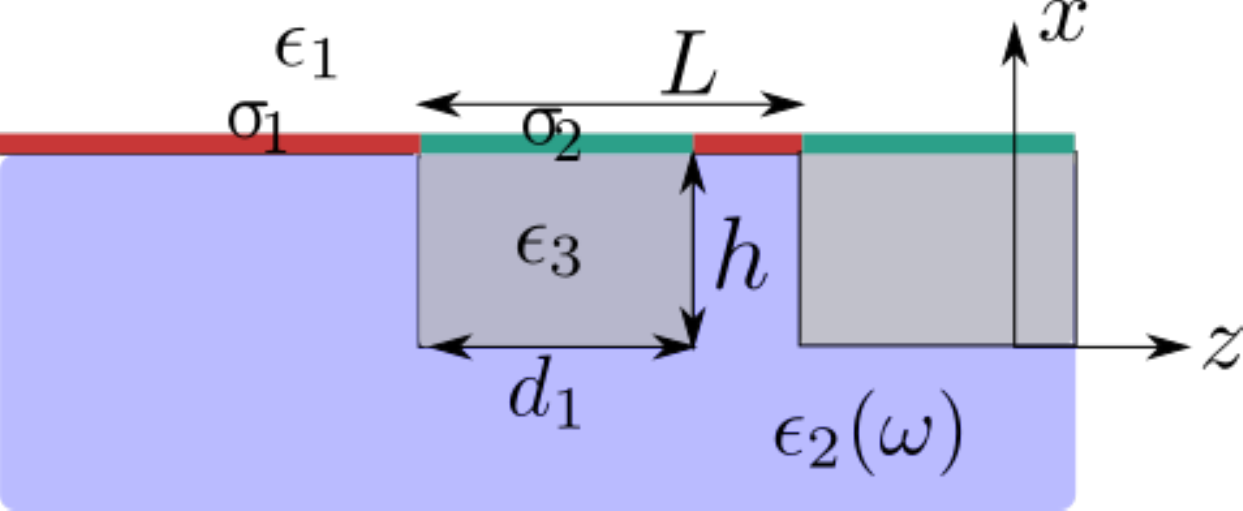}\caption{Unit cell of a dielectric grating. 
  The bottom substrate has dielectric
function $\epsilon_{2}(\omega)<0$, the trench has dielectric function
$\epsilon_{3}$, and the top dielectric has dielectric function $\epsilon_{1}$. Fig. \ref{fig:Unit-cell-of} with the inclusion of
alternating graphene strips with different conductivities $\sigma_1$ and $\sigma_2$. 
We assume that
 a surface-plasmon polariton is impinging from the left on the Bragg
 grating and is scattered from it. The reflectance coefficient, ${\cal R}$,
 is computed using coupled-mode theory.
 \label{fig:Unit-cell-of_graphene} \label{fig:Unit-cell-of}}
\end{figure}

\section{Coupled-mode theory equations} \label{sec:simple}

The reader interested in a detailed derivation of the coupled-mode theory is invited to read the supplementary information. Here, we study a metal-dielectric or dielectric-dielectric
interface, where a Bragg grating, extending over a finite region,
is written on the surface of the metal/dielectric. In addition, we also study the case  of alternating graphene strips with
different conductivities  deposited a the surface, as can be seen in Fig. \ref{fig:Unit-cell-of}.

For solving the scattering problem, we firstly  solve the  unpatterned waveguide problem, i.e., with translation symmetry along the $z$ axis. From this we obtain the 
corresponding electromagnetic normalized modes (see supplementary material)  $\bm{{\cal E}}_{\nu,t}, \bm{{\cal H}}_{\nu,t}$ such that the 
power
per unit length transported along the $z-$direction is  ${\cal P}$.  Afterwards, we  decompose the propagating field inside the patterned heterostructure
as function of the eigenmodes of the unperturbed waveguide:
\begin{eqnarray}
\mathbf{E}_{t} & =\sum_{\mu}a_{\mu}(z)\bm{{\cal E}}_{\mu,t}, \label{eq:E1}\\
\mathbf{H}_{t} & =\sum_{\mu}b_{\mu}(z)\bm{{\cal H}}_{\mu,t}, \label{eq:H1}
\end{eqnarray}
where the sum over $\mu$ has  implicit  summation and integration
over both discrete and continuous modes, respectively.  Each mode coefficient
$a_{\mu}(z)$, $b_\mu(z)$ can be decomposed as function of right and left propagating
coefficients:
\begin{equation}
b_{\mu}(z)=b_{\mu}^{+}(z)e^{i\beta_{\mu}z}-b_{\bar{\mu}}^{-}(z)e^{-i\beta_{\mu}z}, \label{eq:b}
\end{equation}
and 
\begin{equation}
a_{\mu}(z)=b_{\mu}^{+}(z)e^{i\beta_{\mu}z}+b_{\bar{\mu}}^{-}(z)e^{-i\beta_{\mu}z}. \label{eq:a}
\end{equation}

Substituting back Eqs. (\ref{eq:E1}--\ref{eq:a}) in the Maxwell equations, we obtain
 the following coupled-mode equations:
\begin{equation}
\frac{db_{\mu}^{+}}{dz}-i\beta_{\mu}b_{\mu}^{+}=\sum_{\nu}K_{\mu,\nu}^{++}b_{\nu}^{+}+K_{\mu,\nu}^{+-}b_{\nu}^{-}. \label{eq:mode1}
\end{equation}
\begin{equation}
\frac{db_{\mu}^{-}}{dz}+i\beta_{\mu}b_{\mu}^{-}=\sum_{\nu}K_{\mu,\nu}^{-+}b_{\nu}^{+}+K_{\mu,\nu}^{--}b_{\nu}^{-},\label{eq:mode2}
\end{equation}
where the coupling coefficients are
\begin{equation}
K_{\mu,\nu}^{s_1s_2}=s_1K_{\mu,\nu}(z)+s_2k_{\mu,\nu}(z), \label{eq:K_munu_s1s2}
\end{equation}
where $s_1,s_2=\pm1.$ and
\begin{eqnarray}
K_{\mu,\nu}(z) & = & \frac{i\omega\epsilon_{0}}{4{\cal P}}\int dx[\epsilon(x,z)-\epsilon(x)]{\cal E}_{\mu,x}^{\ast}{\cal E}_{\nu,x}
,\label{eq:K_mu_nu}\\
k_{\mu,\nu}(z) & = &
- \frac{ \sigma_1}{\sigma_2(z)}  \frac{\left[\sigma_2(z)-\sigma_1\right] }{4{\cal P}} {\cal E}_{\mu,z}^{\ast}(0){\cal E}_{\nu,z}(0)+ \nonumber\\ &+&
\frac{i\omega\epsilon_{0}}{4{\cal P}}\int dx\frac{\epsilon(x) }{\epsilon(x,z) } \left(\epsilon(x,z) -\epsilon(x)
 \right){\cal E}_{\mu,z}^{\ast}{\cal E}_{\nu,z}.\label{eq:k_mu_nu}
\end{eqnarray}
Here, the  power 
per unit length transported along the $z-$direction reads
\begin{equation}
{\cal P}  =\frac{1}{2}\int dx\Re(\bm{{\cal E}}_{\nu}\times\bm{{\cal H}}_{\nu}^{\ast})\cdot\mathbf{e}_{z}=\frac{\beta}{2\omega\epsilon_{0}}\int dx\frac{1}{\epsilon(x)}{\cal H}_{\nu,y}{\cal H}_{\nu,y}^{\ast}.
\end{equation}

Equations   (\ref{eq:mode1}) and  (\ref{eq:mode2}) describe the propagation of a plasmonic wave along a heterostructure  containing
graphene layers.  The specific geometry and dielectric information of the patterned heterostructure
is encoded into the coupling coefficients $K_{\mu,\nu}(z)$ and $k_{\mu,\nu}(z)$. For the specific case of a square-wave grating (Bragg grating), the integrals in (\ref{eq:K_mu_nu}) and (\ref{eq:k_mu_nu}) are analytical and they will be given in the following sections.
  As we will see in the next sections, for a square-wave grating
it is possible to find an exact analytical solution of the system of equations (\ref{eq:mode1}) and (\ref{eq:mode2}).

\section{Solution for a square-wave Bragg grating} \label{sec:exact_sol}

For a Bragg grating described by alternating dielectrics (where
inside each dielectric slab the permittivity depends only on the 
transverse direction $x$) 
an exact solution can be obtained when only two modes are involved,  a situation appropriate to our case. Let us consider:
\begin{equation}
\epsilon(x,z) = \left\{  \begin{array}{r@{\quad}cr} 
\tilde{\epsilon}(x), \,\,\mathrm{if } \,\,nL<z<d_1+nL,\\ 
\epsilon(x), \,\, \mathrm{if}  \,\,d_1+nL<z<(n+1)L, 
\end{array}\right. \label{eq:lattice}
\end{equation}
with $n$ an integer. The Bragg lattice has $N$ unit cells. In the lattice described by Eq. (\ref{eq:lattice}), the coefficient $K^{pq}_{\mu\nu}$ will be constant inside each slab. From now on we will consider only two modes, the forward and back-scattered with label $\mu$. 
We simplify the notation so $b^+_\mu\equiv X$ and $b^-_\mu \equiv Y$. We define $K_{\mu\mu}^{++}=-K_{\mu\mu}^{--}\equiv u$ and $K^{+-}_{\mu\mu}=-K^{-+}_{\mu\mu}=-v$ when $nL<z<d_1+nL$. Therefore, Eqs. (\ref{eq:mode1}) and (\ref{eq:mode2}) can be written as a set of coupled differential equations:
\begin{eqnarray}
X^\prime-i\beta X=uX +v Y, \label{eq:X}\\
Y^\prime+i\beta Y=-vX-uY. \label{eq:Y}
\end{eqnarray}

The transmission and reflection coefficients can be defined as usual:
\begin{eqnarray}
{\cal R}&=& \left| \frac{Y(0)}{X(0)}\right|^2, \\
{\cal T}&=& \left| \frac{X(NL)}{X(0)}\right|^2.
\end{eqnarray}

Next, we  solve Eqs. (\ref{eq:X}) and (\ref{eq:Y}) for $0<z<d_1$. From (\ref{eq:Y}) we obtain:
\begin{equation}
X=-\frac{Y^\prime+(i\beta+u)Y}{v}, \label{eq:X2}
\end{equation}
which can be combined into a single equation for $Y$ using Eq. (\ref{eq:X}):
\begin{equation}
Y^{\prime\prime}+(v^2+\beta^2-2i\beta u-u^2)Y=0, \label{eq:mhs}
\end{equation}
and defining:
\begin{equation}
g=\sqrt{v^2+(\beta-i u)^2}, \label{eq:g}
\end{equation}
the solution of Eq. (\ref{eq:mhs}) can be written as:
\begin{equation}
Y(z)=Y_+ e^{igz}+Y_-e^{-igz}, \label{eq:Yz}
\end{equation}
with $Y_+$ and $Y_-$ constants.  Substituting back in Eq. (\ref{eq:X2}) we obtain:
\begin{equation}
X(z)=-Y_+ \frac{ig+i\beta+u}{v}^{igz}-Y_-\frac{-ig+i\beta+u}{v}e^{-igz}. \label{eq:Xz}
\end{equation}

For obtaining the transfer matrix of the propagation between $z=0$ and $z=d_1$, 
we need to write the components $Y(z=d_1)$ and $X(z=d_1)$ as function of 
$Y(z=0)$ and $X(z=0)$. Therefore, using Eqs. (\ref{eq:Yz}) and (\ref{eq:Xz}) for $z=0$ we 
obtain:
\begin{subequations}
\begin{eqnarray}
Y(0)&=Y_++Y_- \\
X(0)&=h_+ Y_+ +h_- Y_-,
\end{eqnarray}\label{eq:XY} 
\end{subequations}
where we have defined $h_\pm=-\frac{\pm g+i\beta+}{v}$. For $z=d_1$ we find: 
\begin{subequations}
\begin{eqnarray}
Y(d_1)=Y_+ e^{igd_1}+Y_-e^{-igd_1},\\
X(d_1)=h_+ Y_+ e^{igd_1}+h_-Y_-e^{-igd_1}.
\end{eqnarray}\label{eq:Xd1Yd1} 
\end{subequations}

From Eqs. (\ref{eq:XY}) we have: 
\begin{eqnarray}
Y_+= \frac{X(0)-h_- Y(0)}{h_+-h_-},\\
Y_-= -\frac{X(0)-h_-Y(0)}{h_+-h_-},\label{eq:Ym} 
\end{eqnarray} 
and using Eqs. (\ref{eq:Ym}) in Eqs. (\ref{eq:Xd1Yd1}) we obtain after some algebra:\begin{subequations}
\begin{flalign}
X(d_1)=\left[ \cos(gd_1)+i\Delta_1\sin(gd_1) \right]X(0)+ \nonumber \\+\Delta_2\sin(gd_1) Y(0),\\
Y(d_1)=-\Delta_2\sin(gd_1) X(0) + \left[ \cos(gd_1)-\right. \nonumber\\ \left. -i\Delta_1\sin(gd_1) \right]Y(0),
\end{flalign}\end{subequations}
where we have defined $\Delta_1\equiv\frac{\beta-iu}{g}$ and $\Delta_2\equiv \frac{v}{g}$, with $\Delta_1^2+\Delta_2^2=1$. This defines the propagation along any unit cell from $z=nL$ to $z=d_1+nL$.
The propagation from $z=d_1+nL$ to $z=(n+1)L$ is given from the solution of:
\begin{eqnarray}
X^\prime-i\beta X=0, \nonumber\\
Y^\prime+i\beta Y=0,
\end{eqnarray}
where the coupling constants $K^{s_1,s_2}_{\mu,\mu}$ vanishes because of the dielectric function (\ref{eq:lattice}).
Therefore, we have the trivial solution: $X(z)=X(d_1)e^{i\beta z}$ and $Y(z)=Y(d_1)e^{-i\beta z}$.
The total transfer matrix of the propagation along an entire unit cell is:
\begin{equation}
M=\begin{pmatrix}
e^{i\theta_2} & 0 \nonumber \\ 0 & e^{-i\theta_2}
\end{pmatrix}
\begin{pmatrix}
\cos\theta_1+i\Delta_1 \sin\theta_1 & \Delta_2\sin\theta_1 \nonumber \\
-\Delta_2\sin\theta_1 & \cos\theta_1-i\Delta_1\sin\theta_1
\end{pmatrix},
\end{equation}
where we defined $\theta_1\equiv gd_1$ and $\theta_2\equiv \beta d_2$, such that the
transfer matrix  relating the right and left propagating fields impinging on each face of a unit cell $[X((n+1)L), Y((n+1)L)]^T=M [X(nL), Y(nL)]^T $ is (the super-index $T$ refers to the transpose operation):
\begin{eqnarray}
M=\begin{pmatrix}
e^{i\theta_2}\left(\cos\theta_1+i\Delta_1 \sin\theta_1\right) & e^{i\theta_2}\Delta_2\sin\theta_1\nonumber \\
-e^{-i\theta_2}\Delta_2\sin\theta_1 & e^{-i\theta_2} \left( \cos\theta_1-i\Delta_1\sin\theta_1\right)
\end{pmatrix}. \label{eq:transfer_matrix}
\end{eqnarray}
From the eigenvalues of the above matrix we can obtain the Bloch phase $\gamma$:
\begin{equation}
\cos\gamma=\cos\theta_1\cos\theta_2-\Delta_1\sin\theta_1\sin\theta_2, \label{eq:bloch}
\end{equation}
and from the Chebyshev identity \cite{Soukoulis} we can obtain the transmission and
reflection coefficients for the propagation along $N$ unit cells:
\begin{eqnarray}
T= \frac{\sin^2\gamma}{\sin^2\gamma+|\Delta_2|^2|\sin\theta_1|^2 \sin^2\left(N\gamma \right)}, \label{eq:trans} \\ 
R=\frac{|\Delta_2|^2|\sin\theta_1|^2 \sin^2\left[N\gamma \right]}{\sin^2\gamma+|\Delta_2|^2|\sin\theta_1|^2 \sin^2\left(N\gamma \right)}. \label{eq:refl}
\end{eqnarray}

Therefore,  we  have obtained analytical formulas for the propagation of two coupled modes.
This formalism will be used in the next two sections to
obtain the propagation properties of SPPs in metallic and graphene
gratings.

\section{SPP scattering from a metallic grating \label{sec:SPP}}

In this section we consider the scattering of a SPP from a Brag grating
whose unit cell is represented in Fig. \ref{fig:Unit-cell-of}. The
dielectric function $\epsilon_{1}$ is vacuum, the dielectric $\epsilon_{2}(\omega)$
represents the optical response of the metal below the plasma frequency,
and $\epsilon_{3}$ is another dielectric, in principle different
from $\epsilon_{1}$ and $\epsilon_{2}(\omega)$.  

\subsection{SPP fields and dispersion relation}

Let us consider an interface between a metal and a dielectric. The
relative dielectric function of the metal is in the spectral range where
$\epsilon_{2}(\omega)<0$ and that of the dielectric is $\epsilon_{1}$,
and is assumed constant. The dispersion relation of a surface-plasmon polariton
at a metallic interface with a dielectric is given by \cite{Peres}
\begin{equation}
q=\frac{\omega}{c}\sqrt{\frac{\epsilon_{1}\epsilon_{2}(\omega)}{\epsilon_{1}+\epsilon_{2}(\omega)}}.\label{eq:ddispersion}
\end{equation}

The field
of the SPP has the form \cite{Peres}
\begin{eqnarray}
\mathbf{E}_{\alpha}(\mathbf{r},t) & =(E_{\alpha,x}\mathbf{e}_{x}+E_{\alpha,z}\mathbf{e}_{z})e^{-\kappa_{\alpha}\vert x\vert}e^{i(qz-\omega t)},\label{eq:E_alpha}\\
\mathbf{B}(\mathbf{r},t) & =B_{y}\mathbf{e}_{y}e^{-\kappa_{\alpha}\vert x\vert}e^{i(qz-\omega t)},\label{eq:B}
\end{eqnarray}
where $\alpha=1,2$ defines the medium where the field is located.
Using Maxwell's equations, we obtain
\begin{eqnarray}
 E_{\alpha,x}&=\frac{q}{\omega\epsilon_{0}\epsilon_{\alpha}}H_{y},\\
E_{\alpha,z}&=-i{\rm sgn}(z)\frac{\kappa_{\alpha}}{\omega\epsilon_{0}\epsilon_{\alpha}}H_{y},\\
\kappa_{\alpha} & =\sqrt{q^{2}-\epsilon_{\alpha}\omega^{2}/c^{2}}.
\end{eqnarray}
We note that $E_{\alpha,x}$ and $H_{y}$ are the transverse
fields, whereas  $E_{\alpha,z}$ is the longitudinal component. 
The usual boundary conditions for the fields at an interface, $E_{1,z}=E_{2,z}$
and $B_{1,y}=B_{2,y}$, lead to the dispersion relation (\ref{eq:ddispersion}).
Note that $\epsilon_{1}$ and $\epsilon_{2}$ must have different
signs for satisfying the first boundary condition. Therefore, an SPP
mode only exists when its frequency is below the plasma frequency.
Indeed, we can show from Eq. (\ref{eq:ddispersion}) that the frequency
region for the existence of the SPP obeys the condition
\begin{equation}
\omega<\frac{\omega_{p}}{\sqrt{\epsilon_{1}+1}},
\end{equation}
where $\omega_{p}$ is the plasma frequency of the metal. The determination
of the magnitude $H_{y}$ follows from the normalization condition
(see the supplementary information for a review), which has
the form
\begin{equation}
\int dx\frac{1}{\epsilon(x)}{\cal H}_{y}(x){\cal H}_{y}^{\ast}(x)={\cal P}\frac{2\omega\epsilon_{0}}{\vert q\vert}, \label{eq:norm}
\end{equation}
or, in the case of the SPP field, 
\begin{equation}
\int_{-\infty}^{0}dx\frac{1}{\epsilon_{2}}e^{2\kappa_{2}x}H_{y}^{2}+\int_{0}^{\infty}dx\frac{1}{\epsilon_{1}}e^{-2\kappa_{1}x}H_{y}^{2}={\cal P}\frac{2\omega\epsilon_{0}}{\vert q\vert},
\end{equation}
which leads to 
\begin{equation}
\Leftrightarrow H_{y}^{2}={\cal P}\frac{4\omega\epsilon_{0}}{\vert q\vert}\frac{\kappa_{1}\kappa_{2}\epsilon_{1}\epsilon_{2}}{\kappa_{1}\epsilon_{1}+\kappa_{2}\epsilon_{2}}.
\end{equation}

\subsection{Dielectric profile}

The system upon which the SPP will scatter is a square dielectric
grating of period $L=d_{1}+d_{2}$. As a function of $x$, the dielectric
profile reads
\begin{equation}
\epsilon(x,z)=
\left\{ \begin{array}{cc}
\epsilon_{3} &{\rm for}\;  nL<z<d_{1}+nL\\
\epsilon_{2}(\omega) &{\rm for}\; nL+d_{1}<z<d_{1}+d_{2}+nL\\
\end{array}\right.
\end{equation}
with $n=0,1,2,\ldots$. 
Once the dielectric profile is known, we can compute the coupling
constants $K_{\mu,\nu}$ and $k_{\mu,\nu}$ using Eqs. (\ref{eq:K_mu_nu})
and (\ref{eq:k_mu_nu}). Since $\epsilon(x,z)=\epsilon(x,z+L)$, the coupling constants are
also periodic. In this case, because the dielectric functions vary
in a step-like manner, they can be computed analytically, reading
\begin{widetext}
\begin{equation}
K_{q,q}(z)=
\left\{ \begin{array}{cc}
i\frac{q\epsilon_{1}(\epsilon_{3}-\epsilon_{2})\kappa_{1}\sinh(h\kappa_{2})e^{-h\kappa_{2}}}{\epsilon_{2}(\epsilon_{1}\kappa_{1}+\epsilon_{2}\kappa_{2})} &{\rm for}\;  nL<z<d_{1}+nL\\
0 & {\rm for}\; nL+d_{1}<z<d_{1}+d_{2}+nL\\
\end{array}\right.
,\label{eq:K_mu_nu_anal}
\end{equation}

and
\begin{equation}
k_{q,q}(z)=
\left\{ \begin{array}{cc}
i\frac{(\epsilon_{3}-\epsilon_{2})\kappa_{1}\kappa_{2}^{2}\sinh(h\kappa_{2})e^{-h\kappa_{2}}}{q(\epsilon_{1}\kappa_{1}+\epsilon_{2}\kappa_{2})} &{\rm for}\;  nL<z<d_{1}+nL\\
0 &{\rm for}\;  nL+d_{1}<z<d_{1}+d_{2}+nL\\
\end{array}\right.,\label{eq:k_mu_nu_anal}
\end{equation}
\end{widetext}
from where the coupling constant $K_{q,q}^{s_1s_2}$ (\ref{eq:K_munu_s1s2}) follows; the
parameter $h$ is the height of the dielectric well/barrier. We note
that the coupling constants $K_{q,q}$ and $k_{q,q}$ are
zero when $h\rightarrow0$ or $\epsilon_{3}\rightarrow\epsilon_{2}$,
which corresponds to the perfect interface. Therefore, the reflectance
coefficient ${\cal R}$ is zero in these cases. 

\subsection{Results for a single barrier}

\begin{figure}[h!]
\centering
\includegraphics[width=8cm]{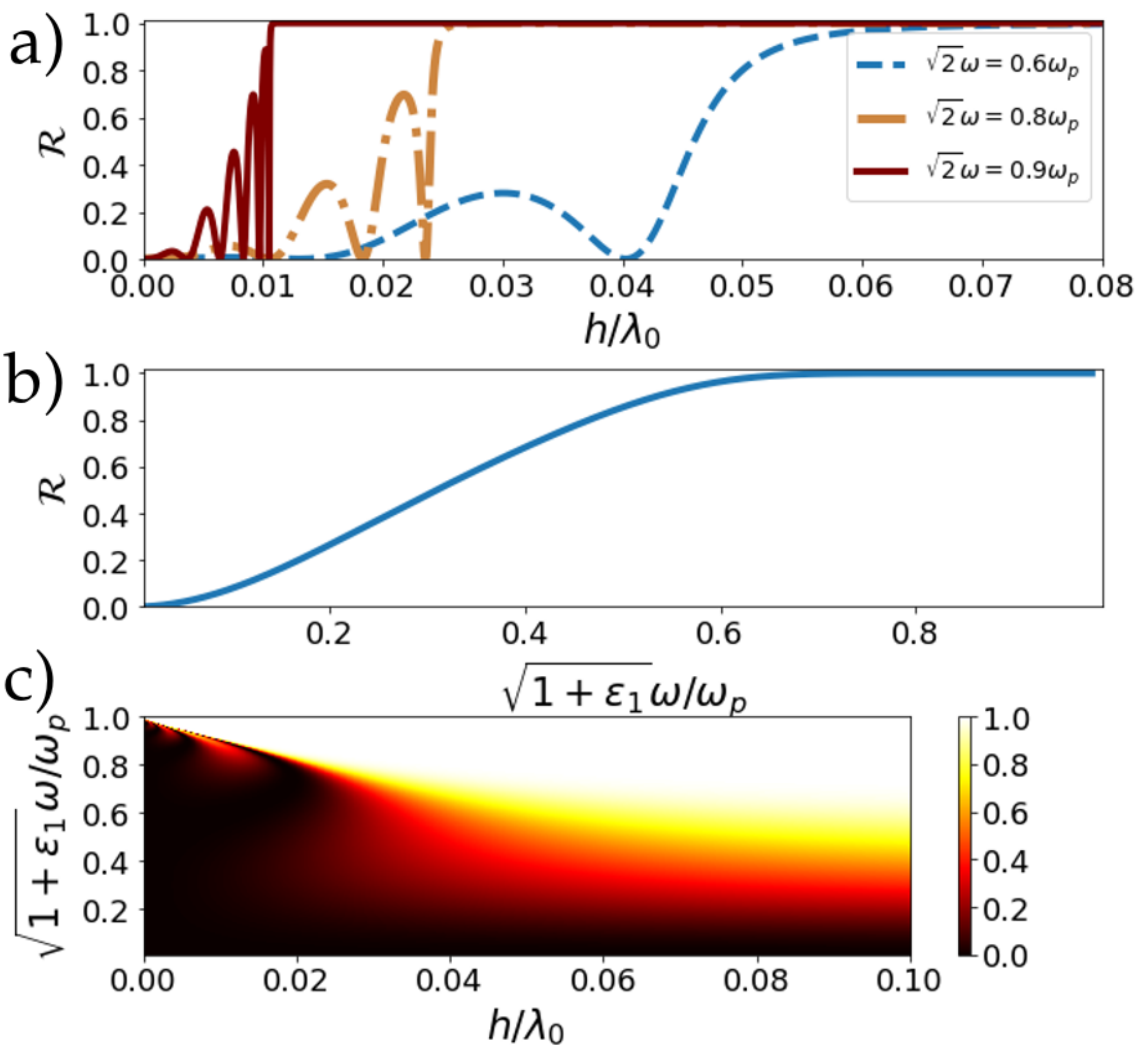} 
\caption{Reflectance coefficient. The plasma frequency was chosen equal to
$\omega_{p}=4$ eV. We take $\epsilon_{3}=\epsilon_1=1$ and $d_1=(1+1/4)\lambda_0$, with
$\lambda_0$ the plasmon wavelength for the frequency $\omega_{spp}=0.6\omega_p/\sqrt{2}$.
(a) Reflectance as function of $h$.
(b) Reflectance as function of the frequency for $h=\lambda_0/2$.  (c) Reflectance as function of
the frequency and $h$.
The dielectric function of the
metal is given by Drude formula $\epsilon_{2}(\omega)=1-\omega_{p}^{2}/\omega^{2}$.\label{fig:Reflectance-coefficient}}
\end{figure}

We first consider the case of a single well of width $d_1$ and height
$h$. The results for the reflectance coefficient are given in Fig.
\ref{fig:Reflectance-coefficient}. In the first panel we see that
${\cal R\rightarrow}0$ when $h\rightarrow0$, as discussed above.
Also there are heights different from zero for which there is
perfect transmission. A similar phenomenon occurs when electrons are
scattered from a potential well. From the central panel of the same
figure we see that ${\cal R}\rightarrow0$ when $\omega\rightarrow0$,
which makes sense since in this case we have essentially free radiation
of very large wavelength and, therefore, unable to see the dielectric
well. As $\omega$ approaches $\omega_{p}/\sqrt{2}$ we have  ${\cal R}\rightarrow1$. The
last panel shows a study of ${\cal R}$ as function of $h$ and $\omega$,
showing that for $h>0.08\lambda_0$ the reflectance becomes insensitive to 
further increases in the well height. The almost absence of oscillations
of the reflectance comes from the fact that $g^2<0$ [see Eq. (\ref{eq:g})] for the parameters used, such that
the fields inside the well are evanescent, as discussed in detail in the next subsection.

\subsection{Results for a Bragg grating}

\begin{figure}
\centering
\includegraphics[width=8cm]{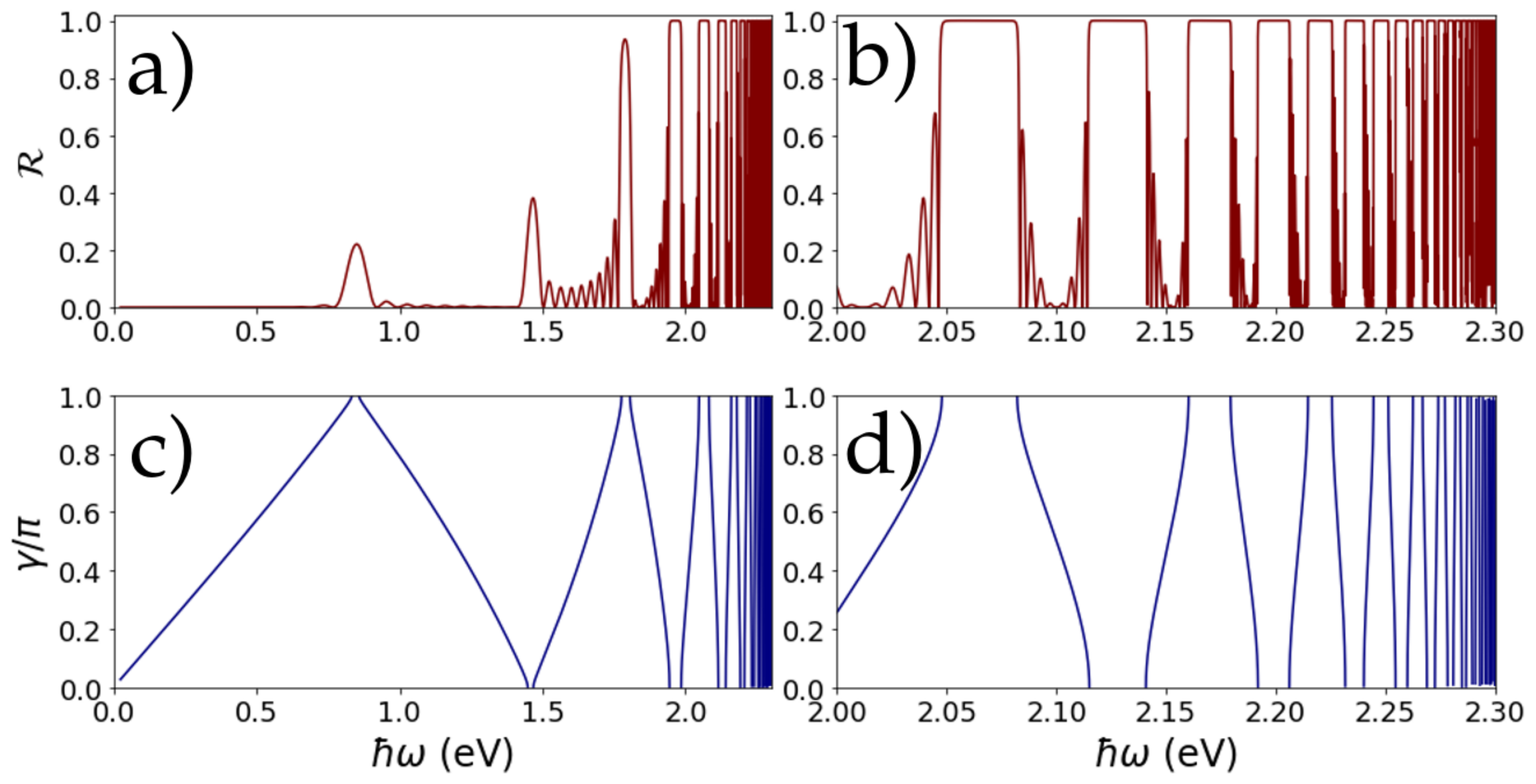}

\caption{ (a), (b) Reflectance of a Bragg grating with $N=10$. We have
chosen $\lambda_0$ the plasmon wavelength for the energy of the SPP given by $\sqrt{2}\omega_{spp}/\omega_{p}=0.6$, $\epsilon_1=1$, $\hbar\omega_p=4$ eV, $d_1=d_2=h=\lambda_0/2$, $\epsilon_3=1-{\omega_p^\prime}^2/\omega^2$,  and $\omega_p^\prime=2.3$ eV.
(c), (d) Dispersion relation as function of the Bloch phase $\gamma=kL$, with $k$ the crystal wavenumber. The bandgaps coincides with the total reflection, as expected. Note the different horizontal scale in the left and right panels.
\label{fig:Reflectance-of-a-Bragg}}

\end{figure}

For a Bragg grating  the    transmission and reflection can be obtained from Eqs. (\ref{eq:trans}) and (\ref{eq:refl}), with 
the Bloch phase given by Eq. (\ref{eq:bloch}). The propagation inside the dielectric well is determined by the value of $g^2$. For $g^2>0$ we have sinusoidal transmission while for $g^2<0$ we have evanescent transmission, with $g^2=0$ determining the crossing between those two regimes. When $\kappa_2h\gg1$ and $\epsilon_3=\epsilon_1$ the function $g^2$ will always be negative, resulting in evanescent transport inside the dielectric well. In this section we will study the
case when the dielectric $3$ is also given by a metal with a corresponding dielectric
function $\epsilon_3=1-{\omega_p^\prime}^2/\omega^2$. 

We show the results for the reflectance and dispersion relation in Fig. \ref{fig:Reflectance-of-a-Bragg}. We consider  $\hbar\omega_p=4$ eV and $\hbar\omega_p^\prime=2.3$ eV. The Bragg grating has $N=10$,  $L=\lambda_0$, $d_1=d_2=L/2$,  where $\lambda_0$ is the wavelength of the plasmon (\ref{eq:ddispersion})  for a frequency of $0.4\omega_p$ and $\hbar\omega_p=4$ eV. We have that $g^2>0$ for $\omega<\omega_p^\prime$ and $g^2\rightarrow[{\omega\to\omega_p^\prime}]{}\infty$. The divergence in $g$ explains the large number of bands slightly 
below the energy of $2.3$ eV in the bottom panel of Fig. \ref{fig:Reflectance-of-a-Bragg}. Note the correlation between the presence of stop-bands in the spectrum of the SPP and the value of 1 for the reflectance.

\section{Scattering of graphene plasmons from a dielectric Bragg grating \label{sec:GSPP}}

Now we consider a system with the geometry of Fig. \ref{fig:Unit-cell-of} but with a graphene sheet on top. Also, we consider that
 conductivity of graphene is $\sigma_1$ ($\sigma_2$) when on top of the dielectric $\varepsilon_2$ ($\varepsilon_3$), as shown in Fig. \ref{fig:Unit-cell-of_graphene} . 
The 
 presence of a graphene sheet corresponds to
including a surface current in the Maxwell equation:
\begin{equation}
\mathbf{\bm{\nabla}\times\mathbf{H}} =\sigma \delta(x)\mathbf{E}_{\parallel}-i\omega\epsilon_{0}\epsilon\mathbf{E}, \label{eq:incl_grap}
\end{equation}
where $\sigma$ is the 2D graphene conductivity  and the sheet is located at the plane $x=0$.

We consider, for simplicity,  that the graphene conductivity is given by the Drude formula \cite{Peres}:
\begin{equation}
\sigma_\mathrm{Drude}= \frac{4i}{\pi}\frac{ E_{F\alpha}\sigma_0}{\hbar\omega+i\Gamma_\alpha}, \label{eq:drude}
\end{equation}
with $\sigma_0=e^2/(4\hbar)$, $E_{F\alpha}$ ($\alpha=1,2$) the Fermi energy in the graphene sheet relative to the Dirac Point, and $\Gamma_\alpha/\hbar$ is
the relaxation rate. From now on we will consider $\Gamma_\alpha=0$ as we are not interested in studying the effect of intrinsic losses. 
In this case,
the boundary condition at the graphene interface leads to a different dispersion relation (when compared to the metallic case) given
by \cite{Peres}
\begin{equation}
\frac{\epsilon_{1}}{\kappa_{1}}+\frac{\epsilon_{2}}{\kappa_{2}}+i\frac{\sigma}{\epsilon_{0}\omega}=0,
\end{equation}
whose solution in the electrostatic limit reads \cite{Peres}
\begin{equation}
q=\frac{\epsilon_{1}+\epsilon_{2}}{4}\frac{(\hbar\omega)^{2}}{\alpha E_{F}\hbar c},\label{eq:dispersion_GSPP}
\end{equation}
where $\alpha\approx1/137$ is the fine structure constant and $E_{F}$
is the Fermi energy of doped graphene. We note that, in addition to
a modification of the dispersion relation, graphene plasmons exist
as well defined excitation for energy scales smaller than the Fermi
energy, and therefore in a different frequency range from that observed
for noble metal plasmonics. 

 We can define a new dielectric tensor taking in account the graphene's conductivity:
\begin{equation}
\tilde{\epsilon}(x)=\epsilon+i\hat{\mathbf{r}}\frac{\sigma\delta(x)}{\omega\epsilon_0}\hat{\mathbf{r}}\cdot, \label{eq:tensor}
\end{equation}
with $\hat{\mathbf{r}}$ the unit vector along the plane that contains the graphene sheet and where the last dot means a inner product when
applied to a vector field.  Therefore Eq. (\ref{eq:incl_grap}) becomes  $\mathbf{\bm{\nabla}\times\mathbf{H}} = -i\omega\epsilon_0\tilde{\epsilon}(x)\mathbf{E}$.

With the presence of a graphene sheet, the tangential 
magnetic field $H_y$ is no longer continuous. Thus, we need to calculate
the new normalization in the sense of Eq. (\ref{eq:norm}),  reading:
\begin{equation}
H_{2,y}^2={\cal P}\frac{4\omega\varepsilon_0}{|q|} \frac{\kappa_1^3\varepsilon_2^2\kappa_2}{\varepsilon_1\kappa_2^3+\varepsilon_2\kappa_1^3}, \label{eq:new_norm}
\end{equation}
and the other components of the electromagnetic field can be calculated from:
\begin{subequations}
\begin{eqnarray}
H_{1,y}=-\frac{\kappa_2 \epsilon_1}{\kappa_1\epsilon_2}H_{2,y}, \\
E_{\alpha,z}=-i\frac{\kappa_\alpha\mathrm{sgn}(x)}{\omega\epsilon_0\epsilon_\alpha}H_{\alpha,y},\\
E_{\alpha,x}=\frac{q}{\omega\epsilon_0\epsilon_\alpha} H_{\alpha,y},
\end{eqnarray} \label{eq:array_gra}
\end{subequations}
with $\alpha=1,2$. Using Eqs. (\ref{eq:new_norm}) and (\ref{eq:array_gra}) in Eqs. (\ref{eq:K_mu_nu}) and (\ref{eq:k_mu_nu}) we obtain:
\begin{equation}
K_{q,q}= iq (\epsilon_3-\epsilon_2) \frac{\kappa_1^3}{\epsilon_2\kappa_1^3+\epsilon_1\kappa_2^3} \sinh(\kappa_2 h) e^{-\kappa_2 h},
\end{equation}
\begin{eqnarray}
k_{q,q}&=&\frac{\sigma_1}{\sigma_2} \frac{\sigma_1-\sigma_2}{\omega\epsilon_0q} \frac{\kappa_1^3\kappa_2^3}{\epsilon_2\kappa_1^3+\epsilon_1\kappa_2^3} +\nonumber \\
&+&i\frac{\epsilon_2}{\epsilon_3}\frac{(\epsilon_3-\epsilon_2) \kappa_2^2\kappa_1^3}{q(\epsilon_2\kappa_1^3+\epsilon_1\kappa_2^3) }\sinh(\kappa_2 h) e^{-\kappa_2 h},
\end{eqnarray}
where $\sigma_\alpha$, with $\alpha=1,2$, is the Drude conductivity (\ref{eq:drude}).

In the following we will show how the formalism works for two examples. First  when $E_{F1}=E_{F2}$ and $\epsilon_2\ne\epsilon_3$. Next we show the results for $E_{F1}\ne E_{F2}$ and $\epsilon_2=\epsilon_3$.

\subsubsection{Different substrates and same conductivity}

\begin{figure}
\centering
\includegraphics[width=8cm]{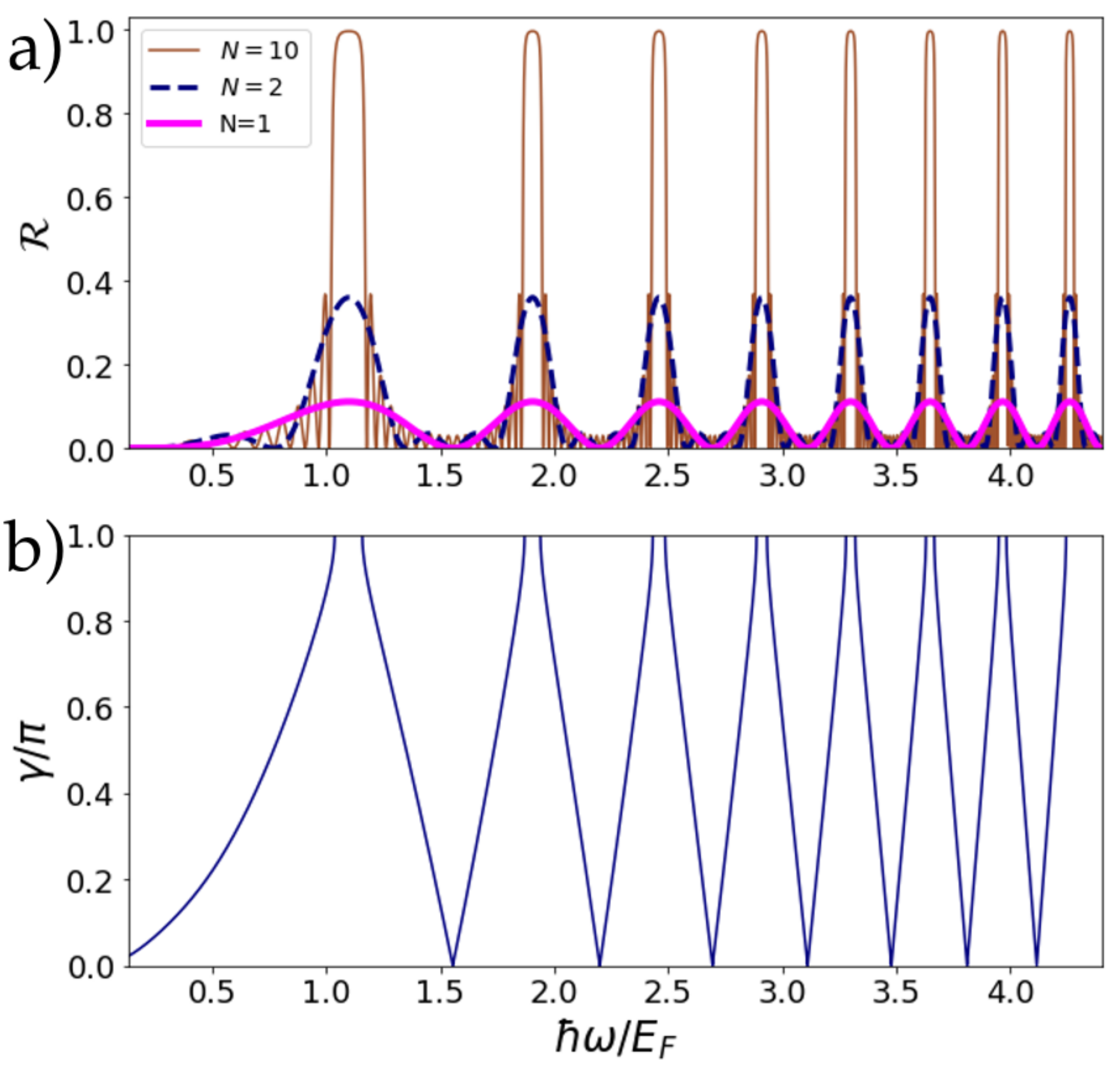}
\caption{(a) Reflectance of a graphene SPP from a Bragg grating with lengths $1L$, $2L$, and $10L$. 
The dielectrics constants are $\epsilon_1=\epsilon_3=1$ and $\epsilon_2=2$. Also we have $E_{F1}=E_{F2}=0.5$ eV. We have fixed the wavelength of the plasmon $\lambda_0$ for the frequency $\hbar\omega_0=1.1E_{F1}$ and we used $h=\lambda_0$, $d_1=\lambda_0/(4\sqrt{F})$, and $d_2=\lambda_0/4$, with $F$ given by Eq. (\ref{eq:def_F}). This choice makes $\theta_1=\theta_2$. We can see a gap opening at the frequency $1.1E_{F1}/\hbar$.
(b) The respective dispersion relation of the plasmonic crystal.
\label{fig:Reflectance-of-a-graphene}}

\end{figure}

\begin{figure}
\centering
\includegraphics[width=8cm]{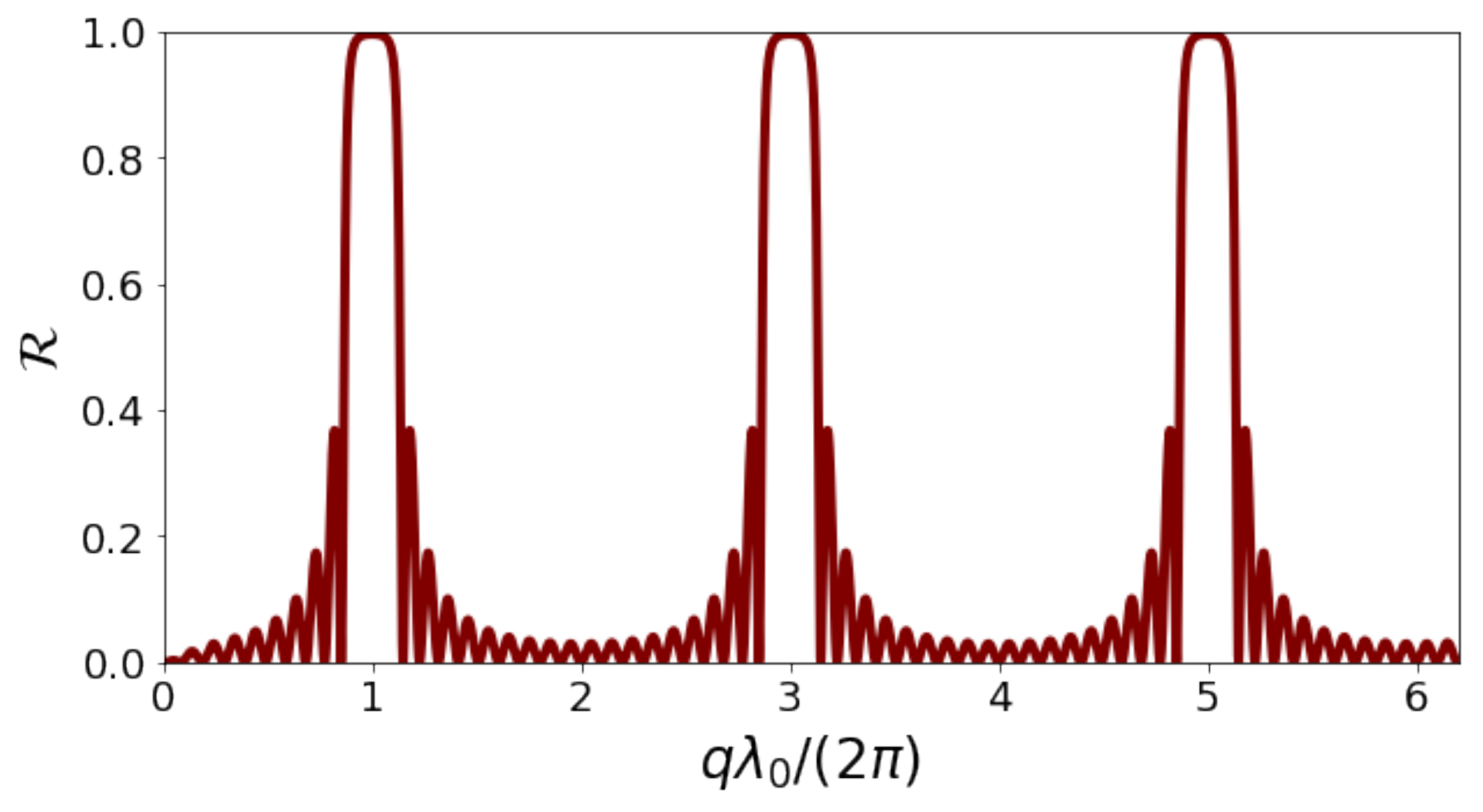}
\caption{Periodicity of the reflection as function of the wavenumber $q$ using the same parameters of Fig. (\ref{fig:Reflectance-of-a-graphene}). The periodicity is a consequence of the phase matching $\theta_1=\theta_2$ between 
the two different cells of the Bragg grating. The stopbands appears when $q\lambda_0=2\pi(1+2n)$, with $n$ an integer. \label{fig:q_dep}}

\end{figure}

Firstly we show the results in Fig. \ref{fig:Reflectance-of-a-graphene} for a grating with $E_{F1}=E_{F2}$ and $\epsilon_3=\epsilon_1=1$ and $\epsilon_2=2$. We can see that the bandwidths decreases as we increase the SPP frequency. In the electrostatic limit and $E_{F1}=E_{F2}$ we can simplify the coefficients $K_{q,q}$, $k_{q,q}$ to:
\begin{equation}
K_{q,q}\rightarrow iq \frac{\epsilon_3-\epsilon_2}{\epsilon_2+\epsilon_1} \frac{\zeta}{2},
\end{equation}
\begin{equation}
k_{q,q}\rightarrow iq \frac{\epsilon_2}{\epsilon_3}\frac{\epsilon_3-\epsilon_2}{\epsilon_2+\epsilon_1} \frac{\zeta}{2},
\end{equation}
with $\zeta=2 \sinh(\kappa_2h)e^{-\kappa_2h}$ and $\zeta\rightarrow1$ when $\kappa_2 h\gg 1$. In this case we have that the functions $K_{qq}^{++}=u$ and $K_{qq}^{-+}=v$ are:
\begin{eqnarray}
u=\zeta\frac{iq}{2} \frac{\epsilon_3^2-\epsilon_2^2}{\epsilon_3(\epsilon_1+\epsilon_2)}, \\
v=-\zeta\frac{iq}{2}\frac{\left( \epsilon_3-\epsilon_2\right)^2}{\epsilon_3(\epsilon_1+\epsilon_2)},
\end{eqnarray}
and we can calculate  $g$ from Eq. (\ref{eq:g}):
\begin{equation}
g=q \sqrt{F(\epsilon_1,\epsilon_2,\epsilon_3,\zeta)},
\end{equation}
with:
\begin{equation}
F=  \frac{(\epsilon_1+(1-\zeta)\epsilon_2+\zeta\epsilon_3)( (1+\zeta)\epsilon_3\epsilon_2+2\epsilon_3\epsilon_1-2\zeta\epsilon_2^2) }{\epsilon_3 (\epsilon_1+\epsilon_2)^2} ,
\label{eq:def_F}
\end{equation}
and thus the functions $\Delta_1=(\beta-iu)/g$ and $\Delta_2=v/g$ that appear inside the transfer matrix  (\ref{eq:transfer_matrix}) are given by:
\begin{eqnarray}
\Delta_1&=&\frac{1}{2} \frac{2\epsilon_3(\epsilon_1+\epsilon_2)\sqrt{F}+ \zeta\epsilon_3^2-\zeta\epsilon_2^2}{\epsilon_3(\epsilon_1+\epsilon_2)\sqrt{F}}, \\
\Delta_2&=&-\zeta\frac{i}{2}\frac{\left( \epsilon_3-\epsilon_2\right)^2}{\epsilon_3(\epsilon_1+\epsilon_2)\sqrt{F}},
\end{eqnarray}
note that as $\omega$ (or $q$) increases we have $\zeta\rightarrow1$. If we ignore the dependence on frequency of the dielectrics constants, the functions $\Delta_1$ and $\Delta_2$ will become frequency independent and also do the propagation properties that are expressed in the transfer matrix approach by Eq. (\ref{eq:trans}). This depends only  on the angles $\theta_1,\theta_2$, both proportional to $q\propto\omega^2$. We show the periodic dependence of the reflection as function of $q$ in Fig. \ref{fig:q_dep}.
In the same figure the stop-band appears for $q=\frac{2\pi}{\lambda_0}(2j+1)$, with $j$ an integer. As before, there is a correlation between the presence of stop-band and the reflectance equal to 1.

\subsubsection{Different conductivities and same substrate}

\begin{figure}
\centering
\includegraphics[width=8cm]{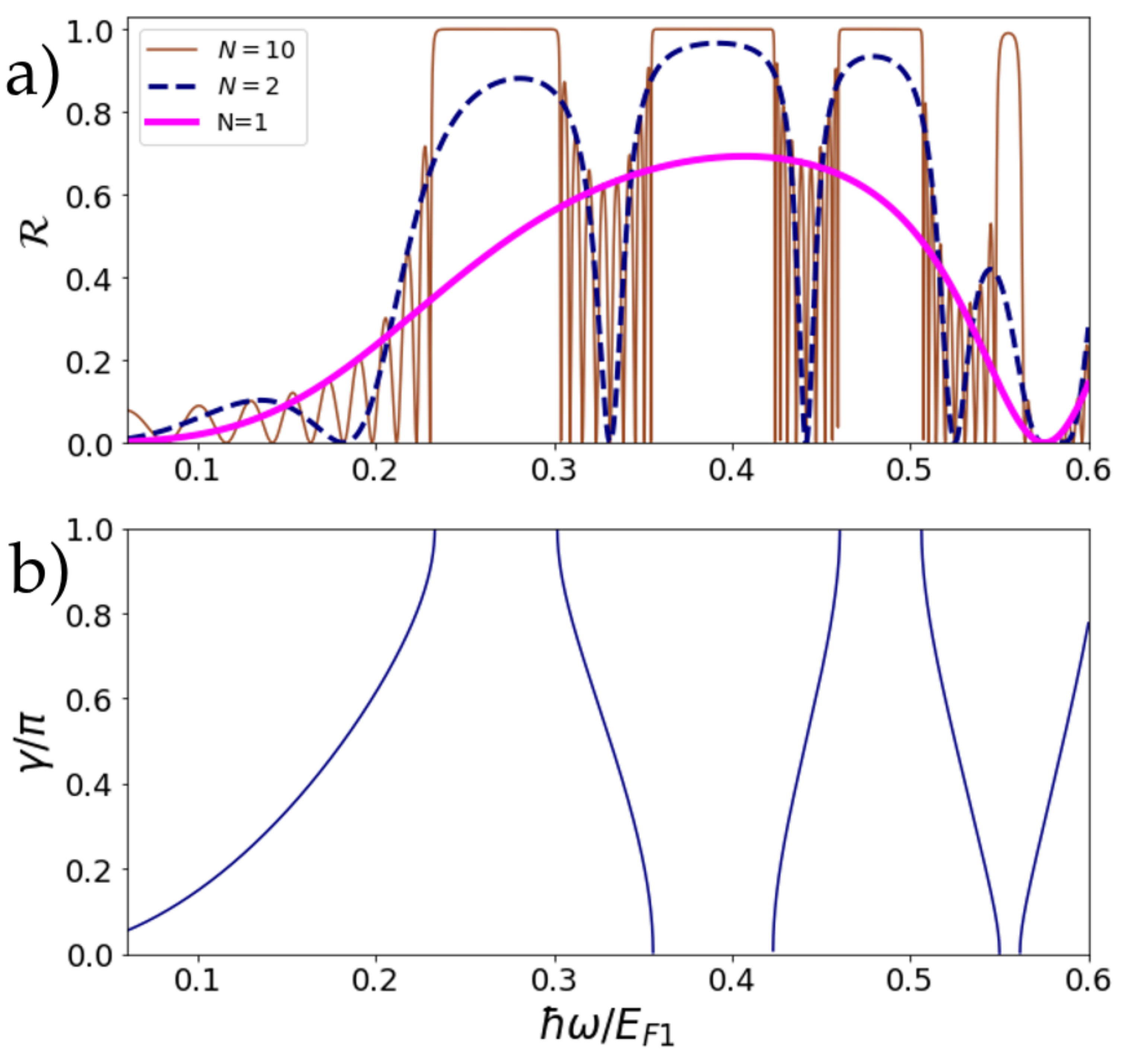}
\caption{ (a) Reflectance of a graphene SPP from a Bragg grating with $N=10$ unit cells. Parameters:
$\epsilon_1=1$, $\epsilon_2=\epsilon_3=4$,$E_{F1}=0.3$ eV, $E_{F2}=0.55$ eV, $\lambda_0$ is the wavelength
for a frequency of $\hbar \omega=E_{F1}/2$ and we fixed $d_1=d_2=\lambda_0/2$.  (b) Dispersion relation for the plasmonic crystal. \label{fig:Reflectance-of-a-graphene2}}

\end{figure}

Next we show in Fig. \ref{fig:Reflectance-of-a-graphene2} the results for a grating of alternating strips of graphene with different Fermi energies deposited on the same substrate with dielectric constant $\epsilon_2$. For the widths we used $d_1=d_2=\lambda_0/2$, where $\lambda_0\approx9\,\mu$m is the graphene plasmon wavelength obtained from Eq. (\ref{eq:dispersion_GSPP}) for $E_{F1}=0.3$ eV and $\hbar\omega=E_{F1}/2$.  We can see that the first stop-band is centered in the frequency $\approx 0.27E_{F1}/\hbar$. Compared to the previous section, we did not impose the phase matching $\theta_1=\theta_2$ condition. Therefore, there is no simple relation between the stop-band frequency and $\lambda_0$ for the chosen parameters. However, we will show in the following that such condition can also be obtained in this case.

In the electrostatic limit  we can obtain $g$, $\Delta_1$, and $\Delta_2$ when $\epsilon_2=\epsilon_3$ as:
\begin{eqnarray}
g^2=2q_1q_2-q_1^2=q_1^2\left(2\frac{E_{F1}}{E_{F2}}-1 \right) \label{eq:g2_graphene},\\
\Delta_1=\frac{q_2}{\sqrt{2q_1q_2-q_1^2}}=\frac{E_{F1}}{\sqrt{2E_{F1}E_{F2}-E_{F2}^2}},\\
\Delta_2=\frac{i(q_2-q_1)}{\sqrt{2q_1q_2-q_1^2}}=\frac{i(E_{F1}-E_{F2})}{\sqrt{2E_{F1}E_{F2}-E_{F2}^2}},
\end{eqnarray}
with $q_i$ obtained using $E_{F}=E_{Fi}$ with $i=1,2$ in Eq. (\ref{eq:dispersion_GSPP}). 

In this limit the functions $\Delta_i$ do not depend on the plasmon frequency and $g=q_1\sqrt{F^\prime}$, with $F^\prime=2E_{F1}/E_{F2}-1$. As we discussed in the previous section, we can also find the condition for the phase matching $\theta_1=\theta_2$:
\begin{equation}
\sqrt{F^\prime}d_1=d_2.
\end{equation}
Thus, we can engineer a stop-band for any given frequency adjusting $d_1,d_2$.  Also, as discussed in the previous subsection, the spectra for $\omega\rightarrow\infty$  becomes periodic with respect to $q\propto\omega^2$, a result that comes from the parameters $\Delta_i$ becoming frequency independent.

\begin{figure}
\centering
\includegraphics[width=8cm]{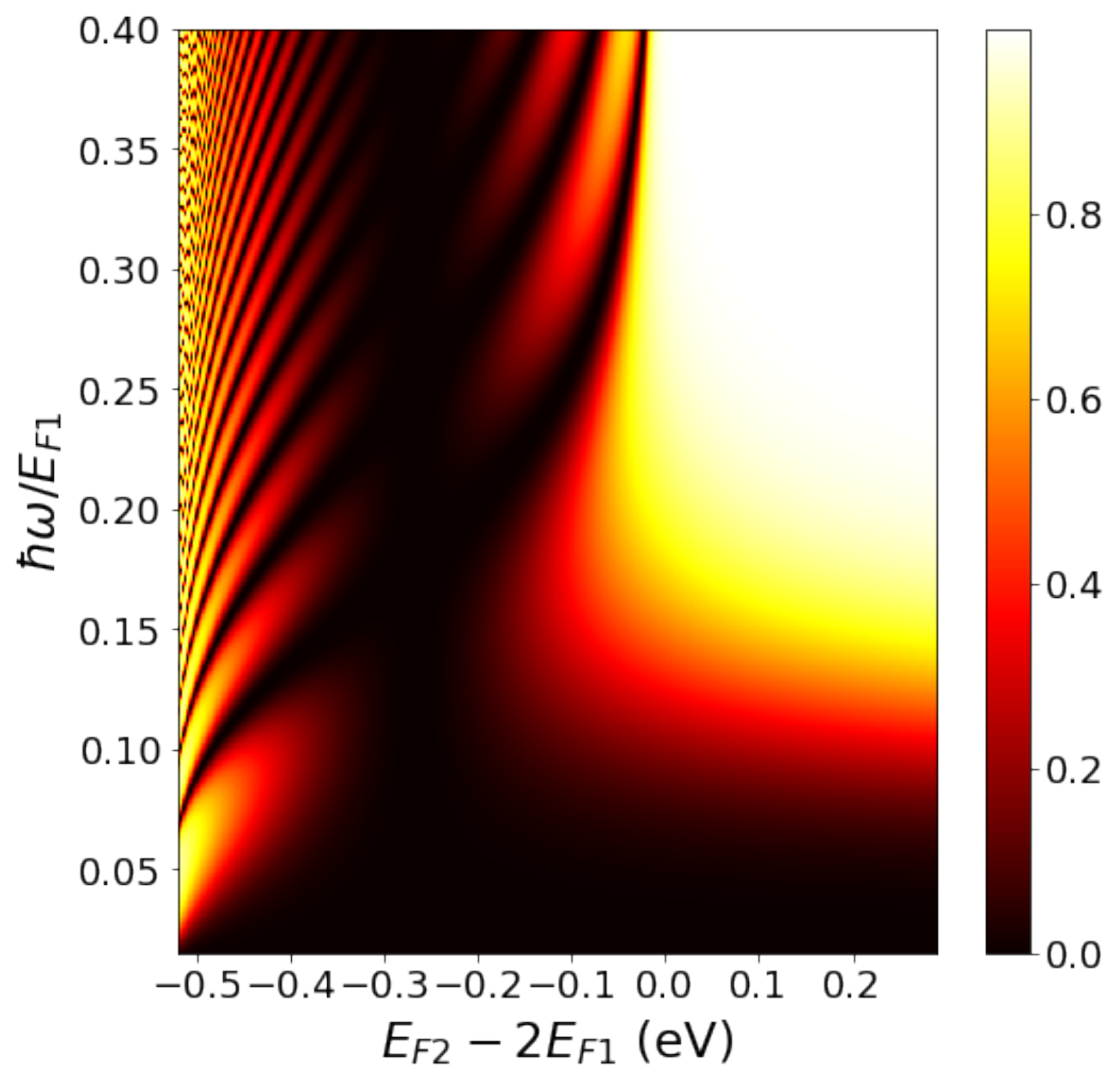}
\caption{Reflectance of a graphene SPP from a single unit cell as 
function of the difference in the Fermi energy and incoming SPP frequency. The bright region
that appears when $E_{F2}-2E_{F1}>0$ is a consequence of the evanescent transport along the graphene
strips in the $\sigma_2$ region. 
The parameters are chosen as: $\epsilon_1=1$, $\epsilon_2=\epsilon_3=4$, $E_{F1}=0.26$ eV. $\lambda_0$ is the wavelength for a plasmon frequency of $0.2E_{F1}/\hbar$ and $d_1=d_2/2=\lambda_0/4$.
\label{fig:Reflectance-of-a-graphene3}}
\end{figure}

In Fig. \ref{fig:Reflectance-of-a-graphene3} we show the reflectance as function of 
$E_{F2}-2E_{F1}$ and the plasmon frequency.  When $E_{F2}-2E_{F1}>0$, we have that $g^2$, given by Eq. (\ref{eq:g2_graphene}), is negative,  implying  evanescence transport and explaining the large bright area with ${\cal R}=1$.

For $E_{F2}\rightarrow0$,
we have that $g/q_1\rightarrow\infty$, meaning that the wavelength in the region with $\sigma_2$ will be 
significantly smaller than the wavelength of the incoming plasmon, making the system acting as a Fabry-Pèrot cavity, explaining the large number of fringes around $E_{F2}-2E_{F1}\approx-0.5$ eV, that is, when $E_{F2}\approx0$. When $E_{F2}\approx E_{F1}$, that happens for $E_{F2}-2E_{F1}=-0.26$ eV and the reflectance goes to zero, as expected.  We can also see that in respect to the difference 
$\Delta E_F=E_{F2}-E_{F1} $ the reflectance is asymmetric. This also happens for an electron scattering through a square well: the reflectance
have different behavior if the square well is positive or negative.

\section{Conclusions\label{sec:conc}}

We have used coupled-mode method and transfer matrix theory for describing the scattering of
a metallic SPP and a graphene SPP from a square-wave Bragg grating
written on the interface between a metal and a dielectric and between
two different dielectrics, respectively. The method allows for simple
analytical expressions for the reflection coefficient ${\cal R}$ and the 
dispersion relation of the Bragg grating.
Our results are valid within the approximations that the coupling of the SPP mode to radiation and evanescent modes
is small.
Relaxing this   approximation  is possible, but analytical
results are no longer available.

We used the analytical results to study the reflection and dispersion relation for different plasmonic Bragg gratings. We characterized the condition for having sinusoidal or evanescent transport. For a Bragg grating consisting of alternating metals with different plasmonic frequencies, we show that there will be an infinite number of bands when the grating metal has a lower plasmonic frequency than the substrate metal.

 We have also shown that the transfer matrix parameters for graphene SPPs are frequency independent in the electrostatic limit. With this result we could find the condition for phase matching of the traveling wave inside each component of the Bragg grating,  thus finding the condition for engineering a stop-band for any given frequency.

Finally, comparing our model with a fully numerical calculation \cite{tao2014graphene} we find the same qualitative behavior, with two small differences: 1) the reflection obtained with our method is larger; 2) the stop band frequency is slightly different. These small differences are due to the consideration of losses in Ref. \cite{tao2014graphene}. On the other hand, it is well known that graphene plasmons  on h-BN have very low losses \cite{woessner2015highly}. Therefore,  if we consider a h-BN buffer layer between graphene and the Bragg grating we expect our model to show a fully quantitative agreement with  numerical solver software.

\section{Suplementary material}

A detailed presentation of coupled mode theory is given in the supplementary information.

\section*{Acknowledgments }

N.M.R.P. acknowledges  Bruno Amorim for discussions
in the early stage of the this work. Both authors thanks D. T. Alves for corrections.  N.M.R.P. acknowledges support
from the European Commission through the project ``Graphene-Driven
Revolutions in ICT and Beyond'' (Ref. No. 785219), COMPETE2020, PORTUGAL2020,
FEDER and the Portuguese Foundation for Science and Technology (FCT)
through project POCI-01-0145-FEDER-028114 and in the framework of
the Strategic Financing UID/FIS/04650/2013.

\end{document}